\documentclass[proceedings]{stacs}
\stacsheading{2010}{669-680}{Nancy, France}
\firstpageno{669}

\usepackage{verbatim}
\usepackage{amssymb}
\usepackage{ifthen}
\usepackage{latexsym}
\usepackage{amstext}

\newboolean{conf}
\setboolean{conf}{true}
\newcommand{\confversion}[1]{\ifthenelse{\boolean{conf}}{#1}{{}}}
\ifthenelse{\boolean{conf}}{}{}

\def \kQBF { \text{\rm QBF}_k }
\def \poly { \text{\rm poly} }
\def \DTS {{\sf DTS}}

\def \P {{\sf P}}
\def \NP {{\sf NP}}

\def \DTISP {{\sf DTISP}}

\def \T {{\sf TIME}}
\def \TIME {{\sf TIME}}
\def \Sig[#1] {{\sf \Sigma}_{#1} }
\def \Pie[#1] {{\sf \Pi}_{#1} }

\def \coNTS {{\sf coNTS}}

\def \SP {{\sf SPACE}}
\def \NT {{\sf NTIME}}
\def \NTIME {\NT}

\def \coNT {{\sf coNTIME}}

\def \DT {{\sf DTIME}}
\def \DTIME {\DT}
\def \isin {\subseteq}
\def \isnotin {\nsubseteq}

\def \eps {\varepsilon}

\begin{document}

\title[Alternation-Trading Proofs, Linear Programming, and Lower Bounds]{Alternation-Trading Proofs, Linear Programming, and Lower Bounds\\(Extended Abstract)}

% \author[ref]{Short author}{Author}
\author[lab1]{R. R. Williams}{Ryan Williams}
% \address[ref]{Address of authors with ref as reference}
\address[lab1]{IBM Almaden Research Center
  \newline 650 Harry Road, San Jose, CA, USA 95120}  %required
\email{ryanwill@us.ibm.com}  %optional
\urladdr{http://www.cs.cmu.edu/~ryanw/}  %optional

\thanks{This material is based on work supported in part by NSF grant CCR-0122581 while the author was a student at Carnegie Mellon University, and NSF grant CCF-0832797 while the author was a member of the Institute for Advanced Study.}

\begin{abstract}

A fertile area of recent research has demonstrated concrete polynomial time lower bounds for solving natural hard problems on restricted computational models. Among these problems are Satisfiability, Vertex Cover, Hamilton Path, $\text{MOD}_6\text{-SAT}$, Majority-of-Majority-SAT, and Tautologies, to name a few. The proofs of these lower bounds  follow a certain proof-by-contradiction strategy that we call {\em alternation-trading}. An important open problem is to determine how powerful such proofs can possibly be.

We propose a methodology for studying these proofs that makes them amenable to both formal analysis and automated theorem proving. We prove that the search for better lower bounds can often be turned into a problem of solving a large series of linear programming instances. Implementing a small-scale theorem prover based on this result, we extract new human-readable time lower bounds for several problems. This framework can also be used to prove concrete limitations on the current techniques.

%Finally, we give an amplifying circuit lower bound for Circuit Evaluation, showing that any $n^{1+\eps}$-size $n^{1-o(1)}$-depth lower bound would already imply ${\sf P}\neq{\sf NC}$. This serves as further evidence that even small polynomial lower bounds may have larger implications than previously believed. %For instance, the current techniques are {\em not} strong enough to prove:
%\begin{itemize}
%\item Satisfiability cannot be solved in $O(n^2)$ time and $n^{o(1)}$ space.

%\item Satisfiability cannot be solved in $O(n^{1.5})$ time on a ``hybrid'' Turing machine, which has random access to its input and sequential access to a worktape.

%\item Tautologies cannot be solved in nondeterministic $O(n^{\phi})$ time and $n^{o(1)}$ space, where $\phi$ is the golden ratio.
%\end{itemize} We conjecture that the above results can be sharpened to match the best lower bounds found by our experiments.

\end{abstract}


\maketitle

%\thispagestyle{empty}

%\newpage

%\setcounter{page}{1}

\section{Introduction}

%This work is concerned with proving new limitations on computers by exploiting their capabilities.
Many known lower bounds for natural problems follow a type of algorithmic argument that we call a {\em resource-trading proof}. Such a proof assumes that a hard problem {\em can} be solved by a ``good'' algorithm, and tries to derive a contradiction by combining two essential components. One is a {\em speedup lemma}, which simulates all good algorithms super-efficiently on some ``interesting'' computational model, trading time for some resource. The second component is a {\em slowdown lemma}, which uses the assumed good algorithm for the hard problem to simulate computations from the ``interesting'' model by good algorithms, thereby trading the ``interesting'' resource for more time. Clever combinations of speedup and slowdown lemmas are used to contradict a known result, in particular some complexity hierarchy theorem. That is, by assuming a ``good'' algorithm for a hard problem, we derive something like $\TIME[n^2] \isin \TIME[n]$, a contradiction.

As an example, one can prove a time-space tradeoff for satisfiability (SAT) as follows. Assume SAT has an algorithm running in $n^c$ time and $\poly(\log n)$ space, for some $c > 1$. One speedup lemma is that computations running in $n^a$ time and $\poly(\log n)$ space can be simulated by an {\em alternating machine} that switches from co-nondeterministic mode to nondeterministic mode once (i.e., a $\Pie[2] $ machine), and runs in $n^{a/2+o(1)}$ time. This speedup lemma {\em trades time for alternations}. The relevant slowdown lemma is: if SAT has an $n^c$ time, $\poly(\log n)$ space algorithm, then (by a strengthening of the Cook-Levin theorem) every language in $\NT[t]$ has $t^{c+o(1)}$ time, $\poly(\log t)$ space algorithms. Consequently, an alternating machine running in $t$ time and making $k-1$ alternations has $t^{c^k + o(1)}$ time, $\poly(\log t)$ space algorithms. Combining these speedup and slowdown lemmas, we derive \[\Sig[2] \T[t] \isin \DTISP[t^{c^2+o(1)},\poly(\log t)] \isin \Pie[2] \T[t^{c^2/2}],\] where the first inclusion holds by slowdown and the second holds by speedup. Now observe that the alternating time hierarchy is contradicted when $c^2 < 2$. This proof is the $n^{\sqrt{2}-\eps}$ time lower bound of Lipton and Viglas~\cite{LV}.

Some of the best known separations in complexity theory use resource-trading proofs. Hopcroft, Paul, and Valiant~\cite{HPV} showed that $\SP[n] \isnotin \DTIME[o(n \log n)]$ for multitape Turing machines, by proving the ``speedup lemma'' that $\DTIME[t] \subseteq \SP[t/\log t]$ and invoking diagonalization. Their result was later extended to general models~\cite{PR,Halpern}. Paul, Pippenger, Szemeredi, and Trotter~\cite{PPST} proved that $\NT[n] \neq \DTIME[n]$ for multitape Turing machines. The key component in the proof is the ``speedup lemma'' $\DT[t] \isin \Sig[4] \T[t/\log^* t]$ for multitape TMs. Despite their age, the above separations still constitute the best known progress on $\P$ vs $\sf PSPACE$ and $\P$ vs $\NP$, respectively.

In more recent years, resource-trading proofs have established time-space lower bounds for ${\sf NP}$-complete problems and problems higher in the polynomial hierarchy~\cite{K84,F,LV,FvM,FLvMV,W06,W08}. For instance, the best known time lower bound for solving SAT with $n^{o(1)}$-space algorithms is $n^{2\cos(\pi/7)-o(1)} \geq n^{1.801}$, obtained with a resource-trading proof~\cite{W08}. (Note if one could improve the $1.801$ exponent to arbitrary constants, one would separate ${\sf LOGSPACE}$ from $\NP$.) %The lower bound also holds for the MOD$m$-SAT problem, for any composite $m$ that is not a prime power. %For complete problems higher up the polytime hierarchy, $k$-QBF (the problem of satisfying quantified Boolean formulas with at most $k$ quantifier blocks) is known to require $\Omega(n^{k+1-\delta_k})$ time for deterministic $n^{o(1)}$ space algorithms~\cite{W07b}, where $\delta_k < 0.2$ rapidly converges towards $0$ as $k$ grows.
For nondeterministic algorithms using $n^{o(1)}$ space, the best known time lower bound for solving the ${\sf coNP}$-complete {\sc Tautology} problem was $n^{\sqrt{2}-o(1)}$ for several years~\cite{FvM}. Certain time-space lower bounds for probabilistic and quantum computations also follow the resource-trading paradigm~\cite{AKRRV,DvM,Viola,vMW}. Resource-trading proofs are also abound in the multidimensional ``hybrid'' Turing machine model, which has read-only random access to its input and an $n^{o(1)}$ read-write store, as well as read-write two-way access to a $d$-dimensional tape for some $d \geq 1$. This is the most powerful (and physically realistic) model known where we still know non-trivial time lower bounds for problems such as SAT. Multidimensional TMs have a long history; e.g., \cite{Loui,PR,K83,MS,vMR,W06} proved lower bounds for them. (For a more complete literature review, please see the full version of the paper.)
%, and the previous best bound for SAT was essentially $O(n^{\sqrt{(d+2)/(d+1)}})$ time for $d$-dimensional tapes.

%\end{itemize}

%For more details please consult the full version of our paper~\cite{full} or Van Melkebeek's recent survey~\cite{VMsurvey2}.

%Resource-trading proofs have been traditionally {\em ad hoc} in their design, making it hard to build intuition about them. One gets a sense that the space of all possible proofs might be difficult to systematically study.

\subsection{Main Results}

We introduce a methodology for reasoning about resource-trading proofs that is also practically implementable for finding short proofs. %We argue that for almost all known resource-trading lower bounds, the proofs can be reformulated in a way that the search for new lower bounds becomes a feasible problem that computers can help attack.\footnote{We note that combinatorial arguments such as Santhanam's time-space lower bound for SAT on multitape Turing machines~\cite{San} do not fall under the alternation-trading paradigm, but they are already known to have different limitations.}
Informally, the ``hard work'' in these proofs can be replaced by solving a series of linear programming problems. This perspective not only aids us practically in the search for new lower bounds, but also allows us to show non-trivial limitations on what can be proved. %These limitations are important, since some components of these proofs do not relativize in some sense (cf. Appendix~\ref{shortintro}).

This methodology is applied to several lower bound problems. In all cases considered here, the resource being ``traded'' is alternations, so we call the proofs {\em alternation-trading}.  %We formalize the components used in prior work and their relevant properties, with the following results.

\noindent{\em Deterministic Time-Space Lower Bounds.} Aided by results of a computer program, we show that any SAT algorithm running in $t(n)$ time and $s(n)$ space satisfies $t \cdot s \geq \Omega(n^{2\cos(\pi/7)-o(1)})$. Previously, the best known result was $t \cdot s \geq \Omega(n^{1.573})$~\cite{FLvMV}. It has been conjectured that the current framework sufficed to prove a $n^{2-o(1)}$ time lower bound for SAT, against algorithms using $n^{o(1)}$ space. %\footnote{An explicit reference for this conjecture , but I have received several referee reports in the past that state it. Also cf.~\cite{LV} in FOCS'99.}
We prove that it is not possible to obtain $n^2$ with the framework, formalizing a conjecture of~\cite{FLvMV}.\footnote{That is, we formalize the statement: ``...some complexity theorists feel that improving the golden ratio exponent beyond 2 would require a breakthrough'' in Section 8 of \cite{FLvMV}.} A computer search over proofs of short length suggests that the best known $n^{2 \cos(\pi/7)-o(1)}$ lower bound~\cite{W08} is already optimal for the framework. We also prove lower bounds on $\kQBF$ (quantified Boolean formulas with at most $k$ quantifier blocks), showing that the problem requires $\Omega(n^{k+1-\delta_k})$ time for $n^{o(1)}$ space algorithms, where $\delta_k < 0.2$ and $\lim_{k \rightarrow \infty} \delta_k = 0$.\footnote{Note the $\kQBF$ results appeared in the author's PhD thesis in 2007 but have been unpublished to date.} %These results appear also optimal for the current tools.

\noindent{\em Nondeterministic Time-Space Lower Bounds.} Adapting our ideas to proving lower bounds for {\sc Tautologies}, a computer program found a very short proof improving upon Fortnow and Van Melkebeek's lower bound. Longer proofs suggested an interesting pattern. Joint work with Diehl and Van Melkebeek on this observation resulted in an $n^{4^{1/3}-o(1)} \geq n^{1.587}$ time lower bound~\cite{DvM2}. Computer search suggests that this lower bound is best possible for the framework. We prove that it is not possible to obtain an $n^{\phi}$ time lower bound, where $\phi = 1.618\ldots$ is the golden ratio. This is surprising since we have known for some time that an $n^{\phi}$ lower bound {\em is} provable for {\em deterministic} algorithms~\cite{FvM}.

\noindent{\em Multidimensional Turing Machine Lower Bounds.} Here our method uncovers peculiar behavior in the best lower bound proofs, regardless of the dimension. Studying computer search results, we extract an $\Omega(n^{r_d})$ time lower bound for the $d$-dimensional case, where $r_d \geq 1$ is the root of a particular quintic $p_d(x)$ with coefficients depending on $d$. For example, $r_1 \approx 1.3009$, $r_2 \approx 1.1887$, and $r_3 \approx 1.1372$. Again, our search suggests this is best possible, and we can prove it is not possible to improve the bound for $d$-dimensional TMs to $n^{1+1/(d+1)}$ with the current tools.

%\smallskip

%\noindent{\bf Size Lower Bounds for Constant Depth Circuits Over General Basis Gates.} Finally, we study a relation between time-space computations and constant-depth circuits recently shown by Allender and Koucky~\cite{AK}. We give lower bounds on the size of uniform circuits of constant depth, where each basis gate is computable in $\DTISP[n,n^{o(1)}]$. (Clearly $\MOD_m$ gates and ${\sf MAJ}$ gates satisfy this condition.) Guided by the theorem prover, we show a size-depth tradeoff lower bound in this setting: SAT requires $n^{1.262}$-size depth-3 circuits, and $n^{1.203}$-size depth-4 circuits. See Appendix~\ref{lowdepth} for a description of these results.

These limitations also hold for other ${\sf NP}$ and ${\sf coNP}$-hard problems; the only property required is that all languages in $\NT[n]$ (respectively, $\coNT[n]$) have sufficiently efficient reductions to the problem. Also our linear programming approach is not limited to the above, and can be applied to the league of lower bounds discussed in Van Melkebeek's surveys~\cite{VMsurvey,VMsurvey2}. %We have chosen to present these cases because in our opinion they are among the most interesting, and the results illustrate a diversity of structure in alternation-trading proofs.

\subsection{Some Remarks on the Reduction to Linear Programming}

The key to our formulation is to separate the {\em discrete choices} in an alternation-trading proof from the {\em real-valued choices}. The discrete choices consist of the sequence of lemmas to apply in each step, and what sort of hierarchy theorem to use in the contradiction. We present several simplifications that greatly reduce the number of discrete choices, without loss of generality. The real-valued choices are the running time exponents that arise from the choices of time bounds and rule applications. We prove that once the discrete choices are made, the remaining real-valued problem can be expressed as an instance of linear programming. This makes it possible to search for new proofs via computer, and it also gives us a formal handle on the limitations of these proofs.

One cannot easily search over all possible proofs, as the number of discrete choices is still about $2^n/n^{3/2}$ for proofs of $n$ lines (proportional to the $n$th Catalan number). Nevertheless it is still feasible to try all $24+$ line proofs. These proof searches reveal patterns, indicating that certain strategies will be most successful in proving lower bounds; in each case we study, the resulting strategies differ. Following the strategies, we establish new lower bound proofs. The patterns also suggest how to show limitations on the proof systems.

\noindent{\bf Note: }{\em Due to space limitations, we can only describe how our methods apply to SAT time-space lower bounds. Please see the full version of the paper for proofs and more details.}

%\noindent{\bf Important Note:} {\em  In the first 12 pages, we can only briefly describe the results and techniques. Please see the Appendices for background information and more details.}

\section{Preliminaries}

We assume familiarity with Complexity Theory, especially the notion of alternation%~\cite{CKS}.
We use big-$\Omega$ notation in the infinitely often sense, so statements like ``SAT is not in $O(n^c)$ time'' are equivalent to ``SAT requires $\Omega(n^c)$ time.'' All functions are assumed constructible within the appropriate bounds. Our default computational model is the random access machine, broadly construed: particular variants do not affect the results. $\DTISP[t(n),s(n)]$ is the class of languages accepted by a RAM running in $t(n)$ time and $s(n)$ space, simultaneously. For convenience, we set $\DTS[t(n)] := \DTISP[t(n)^{1+o(1)},n^{o(1)}]$ to omit negligible $o(1)$ factors.

In order to properly formalize alternation-trading proofs, we introduce notation for alternating complexity classes that include {\em input constraints} between alternations. Let us start with an example of the notation, then give a general definition. Define $(\exists~f(n))^b \DTS[n^a]$ to be the class of languages recognized by a machine which, on an input $x$ of length $n$, writes a $f(n)^{1+o(1)}$ bit string $y$ nondeterministically, copies at most $n^{b+o(1)}$ bits $z$ from the pair $\langle x, y\rangle$ (in $O(n^{b+o(1)})$ time), then feeds $z$ as input to a machine $M$ running in $n^{a+o(1)}$ time and $n^{o(1)}$ space. %We refer to this behavior by saying that {\em $\DTS[n^a]$ is constrained to $n^b$ input}.
Note the runtime of $M$ is measured with respect to the initial input length $n$, not the latter input length $n^{b+o(1)}$ of $z$.

We generalize this definition as follows. Let $\mathcal{C}$ be a complexity class. For $i=1,\ldots,k$, let $Q_i \in \{\exists,\forall\}$ and $a_i, b_i \geq 0$. Define \[(Q_1~n^{a_1})^{b_2}(Q_2~ n^{a_2}) \cdots^{b_k}(Q_k~ n^{a_k})^{b_{k+1}}\mathcal{C}\] to be the class of languages recognized by a machine $M$ that, on input $x$ of length $n$, has the following general behavior on input $x$:

\begin{center}
\begin{tabular}{|l|}\hline
Set $z_0 := x$.\\
For $i=1,\ldots,k$,\\
\hspace{.25in}  If $Q_i = \exists$, switch to {\em existential mode}.\\
\hspace{.25in}  If $Q_i = \forall$, switch to {\em universal mode}.\\
\hspace{.25in}  Guess an $n^{a_i+o(1)}$ bit string $y$ (universally or existentially).\\
\hspace{.25in}  Copy at most $n^{b_{i+1}+o(1)}$ bits $z_i$ from the pair $\langle z_{i-1},y \rangle$.\\
End for\\
Run a machine recognizing a language in class $\mathcal{C}$ on the input $z_k$.\\
\hline
\end{tabular}
\end{center}

When an input constraint $b_i$ is unspecified, its default value is $\max\{a_i,1\}$. We say that the existential and universal modes of an alternating computation are {\em quantifier blocks}, to reflect the complexity class notation. It is crucial to observe that the time bound in the $i$th quantifier block is measured with respect to $n$, the input to the {\em first quantifier block}.

Notice that by simple properties of nondeterminism and conondeterminism, we can combine adjacent quantifier blocks that are of the same type, e.g., $(\exists n^a)^a(\exists n^b)^b\DTS[n^c] = (\exists n^{\max\{a,b\}})^{b}\DTS[n^c]$. This useful property is exploited in alternation-trading proofs.
%writes a $f(n)^{1+o(1)}$ bit string $y$ nondeterministically, copies at most $n^{b+o(1)}$ bits $z$ of the tuple $\langle x, y\rangle$ deterministically (in $O(n^{b+o(1)})$ time), then feeds $z$ as input to a machine recognizing a language from class $\mathcal{C}$. We refer to this behavior by saying that the {\em class $\mathcal{C}$ is constrained to $n^b$ input}. Define $(\exists~ f(n))\mathcal{C} := (\exists~ f(n))^{\max\{1, (\log f(n))/(\log n)\}}\mathcal{C}$. That is, the default input length is assumed to be $f(n)^{1+o(1)}+n^{1+o(1)}$. The class $(\forall~f(n))^b \mathcal{C}$ is defined similarly (with co-nondeterminism).
%Hence a representative machine from the class \[(Q_1~n^{a_1})^{b_2}(Q_2~ n^{a_2}) \cdots^{b_k}(Q_k~ n^{a_k})^{b_{k+1}}\DTS[n^{a_{k+1}}]\] with $Q_i \in \{\exists, \forall\}$ has an input of length $n$ to its first quantifier block, the input to the computation starting at the $i$th quantifier block is of length $n^{b_i+o(1)}$ for $i=2,\ldots,k$, the $i$th quantifier block guesses $n^{a_i}$ bits, the input to the $\DTS$ computation has length $n^{b_{k+1}+o(1)}$, and the final $\DTS$ computation runs in $n^{a_{k+1}}$ time.
%, not to the input to the $i$th quantifier block. %It is important to keep track of the input lengths to quantifier blocks, since several lower bounds rely on the fact that these inputs can be small in certain interesting cases.

\subsection{A Short Introduction to Alternation-Trading Proofs}\label{shortintro}

Here we give a brief overview of the tools used in alternation-trading proofs. In this extended abstract we focus on deterministic time lower bounds for satisfiability for algorithms using $n^{o(1)}$ workspace; the other lower bound problems use similar tools.

It is known that satisfiability of Boolean formulas in conjunctive normal form (SAT) is a complete problem under tight reductions for a small nondeterministic complexity class. The class ${\sf NQL}$, called {\em nondeterministic quasilinear time}, is defined as \[{\sf NQL} := \bigcup_{c \geq 0}\NTIME[n \cdot (\log n)^c] = \NTIME[n \cdot poly(\log n)].\]

\begin{theorem}[\cite{Cook,Schnorr,Tourlakis,FLvMV}]\label{SATfact} SAT is ${\sf NQL}$-complete under quasilinear time $O(\log n)$ space reductions, for both multitape and random access machine models. Moreover, each bit of the reduction can be computed in $O(poly(\log n))$ time and $O(\log n)$ space in both machine models.\footnote{In the multitape Turing machine model we assume that the tape heads are already oriented on the appropriate cells, otherwise it may take linear time to find the appropriate cells on a tape.}
\end{theorem}

Let $\mathcal{C}[t(n)]$ represent a time $t(n)$ complexity class under one of the three models:\begin{itemize}\item deterministic RAM using time $t$ and $t^{o(1)}$ space,\item co-nondeterministic RAM using time $t$ and $t^{o(1)}$ space, \item $d$-dimensional Turing machine using time $t$.\end{itemize}

Theorem~\ref{SATfact} implies that if $\NTIME[n] \isnotin \mathcal{C}[t]$, then SAT~$\notin \mathcal{C}[t/\poly(\log t)]$.

\begin{corollary}\label{satsolve} If $\NTIME[n] \nsubseteq \mathcal{C}[t(n)]$, then SAT$~\notin \mathcal{C}[t(n)/\log^k t(n)]$ for some $k > 0$.
\end{corollary}

Hence we wish to prove $\NT[n] \isnotin \mathcal{C}[n^c]$ for large $c > 1$. To prove time-space lower bounds, we work with $\mathcal{C}[n^c] = \DTS[n^c] = \DTISP[n^c,n^{o(1)}]$.
Van Melkebeek and Raz~\cite{vMR} observed that a similar corollary holds for any problem $\Pi$ such that SAT reduces to $\Pi$ under highly efficient reductions, {\em e.g.} {\sc Vertex Cover}, {\sc Hamilton Path}, {\sc 3-SAT}, and {\sc Max-2-Sat}. Therefore similar time lower bounds hold for these problems as well.

\paragraph{\bf Speedups, Slowdowns, and Contradictions.} Now that our goal is to prove $\NT[n] \isnotin \DTS[n^c]$, how can we do this? In an alternation-trading proof, we attempt to establish a contradiction from assuming $\NT[n] \isin  \DTS[n^c]$, by applying two lemmas which complement one another. A {\em speedup lemma} takes a $\DTS[t]$ class and places it in an alternating class with runtime $o(t)$. A {\em slowdown lemma} takes an alternating class with runtime $t$ and places it in a class with one less alternation and runtime $O(t^c)$. The Speedup Lemma dates back to Nepomnjascii~\cite{Nep} and Kannan~\cite{K84}.

\begin{lemma}[Speedup Lemma]\label{speedup} Let $a \geq 1$, $e \geq 0$ and $0 \leq x \leq a$. Then \[\DTISP[n^a,n^e] \isin (Q_1~n^{x+e})^{\max\{1,x+e\}}(Q_2~\log n)^{\max\{1,e\}}\DTISP[n^{a-x},n^e],\] for $Q_i \in \{\exists,\forall\}$ where $Q_1 \neq Q_2$. In particular, \[\DTS[n^a] \isin (Q_1~n^{x})^{\max\{1,x\}}(Q_2~\log n)^1\DTS[n^{a-x}].\]
\end{lemma}

\proof  Let $M$ use $n^a$ time and $n^e$ space. Let $y$ be an input of length $n$. A complete description (i.e. {\em configuration}) of $M(y)$ at any step can be described in $O(n^e+\log n)$ space. To simulate $M$ in $(\exists~n^{x+e})^{\max\{1,x+e\}}(\forall~\log n)^{\max\{1,e\}}\DTISP[n^{a-x},n^e]$, the algorithm $N(y)$ existentially guesses a sequence of configurations $C_1,\ldots,C_{n^x}$ of $M(x)$. Then $N(y)$ appends the initial configuration $C_0$ of $M(y)$ to the beginning of the sequence, and an accepting configuration $C_{n^x+1}$ to the end. $N(y)$ universally guesses a $i \in \{0,\ldots,n^{x}\}$, erases all configurations except $C_i$ and $C_{i+1}$, then simulates $M(y)$ starting from $C_i$, accepting if and only if $C_{i+1}$ is reached within $n^{a-x}$ steps. It is easy to see the simulation is correct. The input constraints on the quantifier blocks are satisfied since after the universal guess, the input is only $y$, $C_{i}$, and $C_{i+1}$, which is of size $n+2n^{e+o(1)} \leq n^{\max\{1,e\}+o(1)}$. \qed

Observe in the above alternating simulation, the input to the final $\DTISP$ computation is linear in $n+n^e$, {\em regardless of the choice of $x$}. This is a subtle property that is exploited heavily in alternation-trading proofs. The Slowdown Lemma is the following simple result:

\begin{lemma}[Slowdown Lemma] \label{slowdown} Let $a \geq 1$, $e \geq 0$, $a' \geq 0$, and $b \geq 1$. If $\NT[n] \isin \DTISP[n^c,n^e]$, then for both $Q \in \{\exists,\forall\}$, \[(Q~ n^{a'})^{b}\DTIME[n^a] \isin \DTISP[n^{c \cdot \max\{a,a',b\}},n^{e \cdot \max\{a,a',b\}}].\] In particular, if $\NT[n] \isin \DTS[n^c]$, then \[(Q~ n^{a'})^{b}\DTIME[n^a] \isin \DTS[n^{c \cdot \max\{a,a',b\}}].\]\end{lemma}

\proof  Let $L$ be a problem in $(Q~ n^{a'})^{b}\DTIME[n^a]$, and let $A$ be an algorithm recognizing $L$. On an input $x$ of length $n$, $A$ guesses a string $y$ of length $n^{a'+o(1)}$ and feeds an $n^{b+o(1)}$ bit string $z$ to $A'(z)$, where $A'$ is a deterministic algorithm that runs in $n^a$ time. Since $\NT[n] \isin \DTISP[n^c,n^e]$ and $\DTISP$ is closed under complement, by padding we have $\NT[p(n)] \cup \coNT[p(n)] \isin \DTISP[p(n)^c,p(n)^e]$ for polynomials $p(n) \geq n$. Therefore $A$ can be simulated with a deterministic algorithm $B$. Since the runtime of $A$ is $n^{a'+o(1)} + n^{b+o(1)} + n^a$, the runtime of $B$ is $n^{c\cdot \max\{a,a',b\}+o(1)}$ and the space usage is similar.\qed

The final component of an alternation-trading proof is a time hierarchy theorem, the most general of which is the following, provable by a simple diagonalization.

\begin{theorem}[Alternating Time Hierarchy]\label{timehier} For $k \geq 0$, for all $Q_i \in \{\exists, \forall\}$, $1 \leq a'_i < a_i$, and $1 \leq b'_i \leq b_i$, \[(Q_1~n^{a_1})^{b_2} \cdots^{b_k}(Q_k~ n^{a_k})^{b_{k+1}}\DTS[n^{a_{k+1}}] \isnotin (R_1~n^{a'_1})^{b'_2} \cdots^{b'_k}(R_k~ n^{a'_k})^{b'_{k+1}}\DTS[n^{a'_{k+1}}],\] where $R_i \in \{\exists,\forall\}$ and $R_i \neq Q_i$ for all $i=2,\ldots,k+1$.
\end{theorem}

\noindent{\bf Two Examples.} Let us give a couple of examples of alternation-trading proofs. To simplify the presentation we do not specify the input constraints to quantifiers in the below.

(1) In FOCS'99, Lipton and Viglas proved that SAT cannot be solved by algorithms running in $n^{\sqrt{2}-\eps}$ time and $n^{o(1)}$ space, for all $\eps > 0$. By Theorem~\ref{SATfact}, if SAT is in $n^{\sqrt{2}-\eps}$ time and $n^{o(1)}$ space then $\NTIME[n] \isin \DTS[n^c]$ with $c^2 < 2$. We have
\[\begin{array}{lclr}(\exists~n^{2/c^2})(\forall~n^{2/c^2})\DTS[n^{2/c^2}] & \isin & (\exists~n^{2/c^2})\DTS[n^{2/c}] & \text{(Slowdown Lemma)}\\
& \isin & \DTS[n^2] & \text{(Slowdown Lemma)}\\
& \isin & (\forall~n)(\exists~\log n)\DTS[n]. & \text{(Speedup Lemma, with } x = 1\text{)}
\end{array}\]

But $(\exists~n^{2/c^2})(\forall~n^{2/c^2})\DTS[n^{2/c^2}] \isin (\forall~n)(\exists~\log n)\DTS[n]$ contradicts Theorem~\ref{timehier}. In fact, one can show that if $c^2 = 2$, we still have a contradiction with $\NT[n] \isin \DTS[n^c]$, so the $\eps$ can be removed from the previous statement and state that SAT cannot be solved in $n^{\sqrt{2}}$ time and $n^{o(1)}$ exactly.\footnote{Suppose $\NT[n] \isin \DTS[n^c]$ and $\Sig[2] \T[n] \isin \Pie[2] \T[n^{1+o(1)}]$. The first assumption, along with the Speedup and Slowdown Lemmas, implies that for every $k$ there's a $K$ satisfying $\Sig[2] \T[n^k] \isin \NT[n^{kc}] \isin \Sig[K] \T[n]$. But the second assumption implies that $\Sig[K] \T[n] = \Sig[2] \T[n^{1+o(1)}]$. Hence $\Sig[2] \T[n^k] \isin \Sig[2] \T[n^{1+o(1)}]$, which contradicts the time hierarchy for $\Sig[2] \T$.}
\smallskip

(2) Improving on the previous example, one can show SAT $\notin \DTS[n^{1.6004}]$. If $\NTIME[n] \isin \DTS[n^c]$ and $\sqrt{2} \leq c < 2$, then applying the Speedup and Slowdown Lemmas one can derive:
\[\begin{array}{lclr}
\DTS[n^{c^2/2 + 2}] &\isin &(\exists~n^{c^2/2})(\forall~\log n)\DTS[n^2] & \text{(Speedup)}\\
& \isin & (\exists~n^{c^2/2})(\forall~\log n)(\forall~n)(\exists~\log n)\DTS[n] & \text{(Speedup)}\\
& = & (\exists~n^{c^2/2})(\forall~n)(\exists~\log n)\DTS[n] & \text{(Combining $\forall$ Quantifiers)}\\
& \isin & (\exists~n^{c^2/2})(\forall~n)\DTS[n^c] & \text{(Slowdown)}\\
& \isin & (\exists~n^{c^2/2})\DTS[n^{c^2}] & \text{(Slowdown)}\\
& \isin & (\exists~n^{c^2/2})(\exists~n^{c^2/2})(\forall~\log n)\DTS[n^{c^2/2}] & \text{(Speedup)}\\
& = & (\exists~n^{c^2/2})(\forall~\log n)\DTS[n^{c^2/2}] & \text{(Combining $\exists$ Quantifiers)}\\
& \isin & (\exists~n^{c^2/2})\DTS[n^{c^3/2}] & \text{(Slowdown)}\\
& \isin & \DTS[n^{c^4/2}] & \text{(Slowdown)}
\end{array}\]
When $c^2/2 + 2 > c^4/2$ (which happens if $c < 1.6004$), we have $\DTS[n^a] \isin \DTS[n^{a'}]$ for some $a > a'$. %Notice that we do not know if $\DTS[n^a] \isnotin \DTS[n^{a'}]$ when $a' > a$, as the space bounds on both sides of the inequality are the same. However
One can show by a translation argument (similar to the footnote) that either $\DTS[n^a] \isnotin \DTS[n^{a'}]$ or $\NTIME[n] \isnotin \DTS[n^c]$, concluding the proof.
\smallskip

Example (2) was discovered by a computer program. By ``discovered'', we mean that the program applied speedups and slowdowns in precisely the same way, having only minimum knowledge of the lemmas. Furthermore, the program verified that above is the {\em best possible} alternation-trading proof that applies the Speedup and Slowdown Lemmas at most $7$ times. A more formal definition of ``alternation-trading proof'' is given in the next section.

%For readers new to this area, we strongly encourage them to read Appendix~\ref{shortintro}, the {\em Short Introduction to Time-Space Lower Bounds}.

\section{Formalizing Alternation-Trading Proofs}\label{tsSAT}

%We now describe the formalism for alternation-trading proofs of lower bounds on $\NT[n]$ problems, such as SAT. %We shall describe the approach in some detail here; the other settings assume knowledge of this Section.
%Alternation-trading proofs apply a sequence of ``speedup'' and ``slowdown'' lemmas in some order, with the goal of reaching a contradiction by a time hierarchy theorem.
We formalize alternation-trading proofs of lower bounds on $\DTS$ classes as follows:\footnote{This formalization has {\em implicitly} appeared in several prior works, but not to the degree that we investigate in this paper.}

\begin{definition} Let $c > 1$. An {\em alternation-trading proof for $c$} is a list of complexity classes of the form: \begin{equation}\label{simple}(Q_1~ n^{a_1})^{b_2}(Q_2~ n^{a_2}) \cdots^{b_k}(Q_k~n^{a_k})^{b_{k+1}}\DTS[n^{a_{k+1}}],\end{equation} where $k \geq 0$, $Q_i \in \{\exists,\forall\}$, $Q_i \neq Q_{i+1}$, $a_i > 0$, and $b_i \geq 1$, for all $i$. (When $k=0$, the class is deterministic.) The items of the list are called {\em lines of the proof}. Each line is obtained from the previous line by applying either a {\em speedup rule} or a {\em slowdown rule}. More precisely, if the $i$th line is

{\centerline{$(Q_1~ n^{a_1})^{b_2}(Q_2~ n^{a_2}) \cdots^{b_k} (Q_k~ n^{a_k})^{b_{k+1}}\DTS[n^{a_{k+1}}],$}}

then the $(i+1)$st line has one of four possible forms:
 
\noindent {\bf Speedup Rule 0:} For $k = 0$ and any $x \in (0,a_1)$, $(Q_0~ n^{x})^{\max\{x,1\}} (Q_1~n^0)^1\DTS[n^{a_1-x}]$.\footnote{Please note that the $(k+1)$th quantifier is $n^0$ in order to account for the $O(\log n)$ size of the quantifier.}

\noindent {\bf Speedup Rule 1:} For $k > 0$ and any $x \in (0,a_{k+1})$, \[(Q_1~ n^{a_1})^{b_2}(Q_2~ n^{a_2}) \cdots^{b_k} (Q_k~ n^{\max\{a_k,x\}})^{\max\{x,b_{k+1}\}} (Q_{k+1}~n^0)^{b_{k+1}}\DTS[n^{a_{k+1}-x}].\]

\noindent {\bf Speedup Rule 2:} For $k > 0$ and any $x \in (0,a_{k+1})$, \[(Q_1~ n^{a_1})^{b_2} \cdots^{b_k} (Q_k~ n^{a_k})^{b_{k+1}}(Q_{k+1}~ n^{x})^{\max\{x,b_{k+1}\}} (Q_{k+2}~n^{0})^{b_{k+1}}\DTS[n^{a_{k+1}-x}].\]

\noindent {\bf Slowdown Rule:} For $k > 0$, \[(Q_1~ n^{a_1})^{b_2}(Q_2~ n^{a_2}) \cdots^{b_{k-1}} (Q_{k-1}~n^{a_{k-1}})^{b_{k}} \DTS[n^{c \cdot \max\{a_{k+1},a_k,b_k,b_{k+1}\}}].\]

An alternation-trading proof {\em shows} $(\NT[n] \isin \DTS[n^c] \Longrightarrow A_1 \subseteq A_2)$ if its first line is $A_1$ and its last line is $A_2$.
\end{definition}

The above definition comes directly from the Speedup Lemma~(Lemma~\ref{speedup}) and  Slowdown Lemma~(Lemma~\ref{slowdown}). %(Note the $n^0$ in the Speedup Lemma corresponds to $\log n \leq n^{o(1)}$, which is negligible.)
The rules are easily verified to be syntactic formulations of the corresponding lemmas. %, where the $\DTS$ part of the sped-up computation only reads two guessed configurations---so the input it reads is different from the input read by the innermost quantifier block.
For instance, Speedup Rule 1 holds, as
\begin{eqnarray*}& (Q_1~n^{a_1})^{b_2}(Q_2~ n^{a_2}) \cdots^{b_k}(Q_k~ n^{a_k})^{b_{k+1}} \DTS[n^{a_{k+1}}]  &\\ & \isin (Q_1~ n^{a_1})^{b_2}(Q_2~ n^{a_2}) \cdots^{b_k}(Q_k~ n^{a_k})^{b_{k+1}} (Q_k~ n^x)^{\max\{b_{k+1},x\}}(Q_{k+1}~n^0)^{b_{k+1}} \DTS[n^{a_{k+1}}] &\\ & \isin (Q_1~n^{a_1})^{b_2} (Q_2~n^{a_2}) \cdots^{b_k}(Q_k~n^{\max\{a_k,x\}})^{\max\{b_{k+1},x\}}(Q_{k+1}~n^0)^{b_{k+1}} \DTS[n^{a_{k+1}}]. &\end{eqnarray*}
Rule 2 is akin to Rule 1, except that it uses opposite quantifiers in its invocation of the Speedup Lemma.
The Slowdown Rule works analogously to Lemma~\ref{slowdown}.
It follows that alternation-trading proofs are sound.

Note Speedup Rules 0 and 2 add two quantifier blocks, Speedup Rule 1 adds one quantifier, and all three rules introduce a parameter $x$. By considering ``normal form'' proofs (defined in the following paragraphs), we can prove that Rule 2 can always be replaced by applications of Rule 1. (A proof is in the full version of the paper.) For this reason we just refer to {\em the Speedup Rule}, depending on which of Rule 0 or Rule 1 applies.

Define a class of the form (\ref{simple}) to be {\em simple}. %Lower bound proofs using the alternation-trading scheme apply the assumption $\NT[n] \isin \DTS[n^c]$ to derive a contradiction with some time hierarchy. To consider the possible ways of getting a contradiction, we consider the setting of deriving a contradiction from
Define classes $A_1$ and $A_2$ to be {\em complementary} if $A_1$ is the class of complements of languages in $A_2$. Every known (model-independent) time-space lower bound for SAT shows ``$\NT[n] \isin \DTS[n^c]$ implies $A_1 \subseteq A_2$'', for some complementary simple classes $A_1$ and $A_2$, contradicting a time hierarchy (cf. Theorem~\ref{timehier}). A similar claim holds for nondeterministic time-space lower bounds against tautologies (which prove ``$\NT[n] \isin \coNTS[n^c]$ implies $A_1 \subseteq A_2$''), for $d$-dimensional TM lower bounds (which prove ``$\NT[n] \isin \DTIME_d[n^c]$ implies $A_1 \subseteq A_2$''), and other problems.

\noindent{\bf Normal Form.} It will be very convenient to introduce a {\em normal form} for alternation-trading proofs. We show that any lower bound provable with complementary simple classes can also be established with a normal form proof. This greatly reduces the degrees of freedom in a proof, as we no longer need to worry about {\em which} time hierarchy to contradict.

\begin{definition} Let $c \geq 1$. An alternation-trading proof for $c$ is in {\em normal form} if (a) the first and last lines are $\DTS[n^a]$ and $\DTS[n^{a'}]$ respectively, for some $a \geq a'$, and (b) no other lines are $\DTS$ classes.
\end{definition}

We show that a normal form proof for $c$ implies that $\NT[n] \isnotin \DTS[n^c]$.

\begin{lemma}\label{DTShier} Let $c \geq 1$. If there is an alternation-trading proof for $c$ in normal form having at least two lines, then $\NT[n] \isnotin \DTS[n^c]$. \end{lemma}

\begin{theorem} Let $A_1$ and $A_2$ be complementary. If there is an alternation-trading proof $P$ for $c$ that shows $(\NT[n] \isin \DTS[n^c] \Longrightarrow A_1 \isin A_2)$, then there is a normal form proof for $c$, of length at most that of $P$. \label{normalform}
\end{theorem}

Proofs of Lemma~\ref{DTShier} and Theorem~\ref{normalform} are in the full version. The upshot of these results is that we may focus our proof search on normal form proofs. For the remainder of this section, we assume all alternation-trading proofs are in normal form.

\smallskip

\noindent{\bf Proof Annotations.} Different lower bound proofs can result in quite different sequences of speedups and slowdowns. A {\em proof annotation} represents such a sequence.

\begin{definition} A {\em proof annotation} for an alternation-trading proof of $\ell$ lines is the $(\ell-1)$-bit vector $A$ where for all $i = 1,\ldots,\ell-1$, $A[i] = 1$ (respectively, $A[i] = 0$) if the $i$th line applies a Speedup Rule (respectively, a Slowdown Rule).
\end{definition}

An $(\ell-1)$-bit proof annotation corresponds to a ``strategy'' for an $\ell$-line proof. For a normal form proof of $\ell$ lines, it is not hard to show that its annotation $A$ must have $A[1] = 1$, $A[\ell-2] = 0$, and $A[\ell-1]=0$.

Note that an annotation {\em does not} determine a proof entirely, as other parameters need optimizing. (The problem of optimizing them is tackled in the next section.) To illustrate the annotation concept, we give four examples. \begin{itemize}

\item The $n^{\sqrt{2}}$ lower bound of Lipton and Viglas has the annotation $[1,0,0]$.

\item The $n^{1.6004}$ bound from Section~\ref{shortintro} corresponds to $[1,1,0,0,1,0,0]$.

\item The $n^{\phi}$ bound of Fortnow and Van Melkebeek~\cite{FvM} is an inductive proof, corresponding to an infinite sequence of annotations. In normal form, the sequence is $[1,0,0],[1,1,0,0,0],[1,1,1,0,0,0,0],\ldots$

\item The $n^{2 \cos(\pi/7)}$ bound~\cite{W08} has two inductive stages. Let $A = 1,0,1,0,\ldots,1,0,0$, where the `$\ldots$' contain any number of repetitions. The sequence is\\
    \centerline{$[A], [1,A,A], [1, 1,A,A,A], [1,1,1,A,A,A,A],\ldots$}

    That is, the proof performs many speedups, then a sequence of many slowdown-speedup alternations, then two consecutive slowdowns, repeating this until all the quantifiers have been removed.

\end{itemize}

\subsection{Translation To Linear Programming}

Given a (normal form) proof annotation, how can we determine the best proof possible with it? We need to optimally set the runtimes of the first and last $\DTS$ classes in the proof, as well as the $x_i$ parameters that arise from each application of a Speedup Rule. It turns out that an annotation $A$ and $c > 1$ can be reduced to a polynomial size linear program that is feasible if and only if there is an alternation-trading proof of $\NT[n] \isnotin \DTS[n^c]$ with annotation $A$. More precisely, the problem of optimizing parameters can be viewed as an arithmetic circuit evaluation, where the circuit has $\max$ gates, addition gates, and input gates that may multiply their input by $c$. Such circuits can be evaluated using a linear program that minimizes the sum of the gate values (cf.~\cite{Derman}). %Let us give more detail.

Let $A$ be an annotation of $\ell-1$ bits, and let $m$ be the maximum number of quantifier blocks in any line of $A$ (note $m$ is easily computed in linear time). The target LP has variables $a_{i,j}$, $b_{i,j}$, and $x_i$, for all $i = 0,\ldots,\ell-1$ and $j = 1,\ldots,m$. The variables $a_{i,j}$ represent the runtime exponent of the $j$th quantifier block in the class on the $i$th line, $b_{i,j}$ is the input exponent to the $j$th quantifier block of the class on the $i$th line, and for all lines $i$ that use a Speedup Rule, $x_i$ is the choice of $x$ in the Speedup Rule. For example:\begin{itemize}

\item If the $k$th line of a proof is $\DTS[n^a]$, the corresponding constraints are\newline\centerline{$a_{k,1} = a,~~ b_{k,1} = 1,~~(\forall k > 0)~ a_{k,i} = b_{k,i} = 0$.}

\item If the $k$th line of a proof is $(\exists~n^{a'})^{b}\DTS[n^a]$, then the constraints are \newline\centerline{$a_{k,0} = a,~~ b_{k,1} = b,~~a_{k,1} = a',~~b_{k,1} = 1,~~(\forall k > 1)~ a_{k,i} = b_{k,i} = 0.$}\end{itemize}

The objective is to minimize $\sum_{i,j} (a_{i,j} + b_{i,j}) + \sum_i x_i$. The LP constraints depend on the lines of the annotation, as follows. %For every line $i$, we include the constraints $(\forall j:~1\leq j \leq m-1)~ a_{i,j} \leq b_{i,j+1}$.\footnote{These constraints are not strictly necessary, but they are useful in guiding an LP solver, and they are implied by the definition of alternation-trading proof. For more details see the full version of the paper.}

\smallskip

\noindent{\bf Initial Constraints.} For the $0$th and $(\ell-1)$th lines we have $a_{0,1} \geq a_{\ell-1,1}$, and \[a_{0,1} \geq 1,~ b_{0,1} = 1,~~ (\forall~k > 1)~a_{0,k} = b_{0,k} = 0,~~\text{and}~~a_{\ell,1} \geq 1,~ b_{\ell,1} = 1,~~ (\forall k > 1)~a_{\ell,k} = b_{\ell,k} = 0,\] representing $\DTS[n^{a_{0,1}}]$ and $\DTS[n^{a_{\ell-1,0}}]$, respectively. The 1st line of a proof always applies Speedup Rule 1, having the form $(Q_1 n^{x})^{\max\{x,1\}}(Q_2~n^0)^1\DTS[n^{a-x}]$.  So the constraints for the 1st line are: \[\begin{array}{c} a_{1,1} = a_{0,1} - x_1,~b_{1,1} = 1,~ a_{1,2} = 0,~b_{1,2} \geq x_1,~b_{1,2} \geq 1, ~a_{1,3} = x_3,~b_{1,3} = 1,\\(\forall~ k:~ 4 \leq k \leq m)~a_{1,k} = b_{1,k} = 0.\end{array}\]  The below constraint sets simulate the Speedup and Slowdown Rules:

\smallskip

\noindent{\bf Speedup Rule Constraints.} For the $i$th line where $i > 1$ and $A[i] = 1$, we have \[ \begin{array}{c} a_{i,1} \geq 1,~ a_{i,1} \geq a_{i-1,1} - x_i,~b_{i,1} = b_{i-1,1},~a_{i,2} = 0,~b_{i,2} \geq x_i,~b_{i,2} \geq b_{i-1,1},~ a_{i,3} \geq a_{i-1,2},\\~a_{i,3} \geq x_i,~b_{i,3} \geq b_{i-1,2},~(\forall ~k:~4 \leq k \leq m)~a_{i,k} = a_{i-1,k-1}, b_{i,k} = b_{i-1,k-1}.\end{array}\] The constraints express that $\cdots~^{b_2}(Q_2~ n^{a_2})^{b_1}\DTS[n^{a_1}]$ in the $(i-1)$th line is replaced by \[\cdots~^{b_2}(Q_2~ n^{\max\{a_2,x\}})^{\max\{x,b_1\}}(Q_1~n^0)^{b_1}\DTS[n^{\max\{a_1-x,1\}}]\] in the $i$th line, where $Q_1$ is opposite to $Q_2$.

\noindent{\bf Slowdown Rule Constraints.} For the $i$th line where $A[i] = 0$, the constraints are \[\begin{array}{c} a_{i,1} \geq c \cdot a_{i-1,1},~ a_{i,1} \geq c \cdot a_{i-1,2},~ a_{i,1} \geq c \cdot b_{i-1,1},~ a_{i,1} \geq c \cdot b_{i-1,2},~ b_{i,1} = b_{i-1,2}\\ (\forall~k:~2 \leq k \leq m-1)~a_{i,k} = a_{i-1,k+1},~b_{i,k} = b_{i-1,k+1},~a_{i,m} = b_{i,m} = 0.\end{array}\]
These express the replacement of $\cdots~^{b_2}(Q_1 n^{a_2})^{b_1}\DTS[n^{a_1}]$ in the $(i-1)$th line with \[\cdots~^{b_2}\DTS[n^{c \cdot \max\{a_1,a_2,b_1,b_2\}}]\] in the $i$th line.

This concludes the description of the linear program. To find the largest $c$ that still yields a feasible LP, we can simply binary search for it. The following summarizes this section.

\begin{theorem} Given an annotation of $n$ lines, the best possible alternation-trading proof following the annotation can be determined up to $n$ digits of precision, in $\poly(n)$ time.\end{theorem}

\subsection{Results}

Following the above formulation, we wrote proof search routines in Maple. Many millions of proof annotations were tried, including all those corresponding to prior work, with no success beyond the $2 \cos(\pi/7)$ exponent. The best lower bounds followed a highly regular pattern; see the full version for more on this. %For a $424$ line annotation following the pattern, the optimal exponent was only in the interval $[1.80175, 1.8018)$.  One interesting sequence of annotations from the pattern is \[1^k 11000(10)0(10)^20\cdots(10)^k0,\] for $k \geq 2$. One can prove that this sequence cannot yield any lower bound better than $2 \cos(\pi/7)$. %A direct proof of this claim would be extremely technical; using a replacement argument, we can establish the equivalence elegantly.
%\begin{theorem}\label{nobetter2} In the limit (as $k \rightarrow \infty$), the maximum lower bound provable with the sequence of annotations $1^k 11000(10)0(10)^20(10)^30\cdots(10)^k0$ for $k \geq 0$ is that SAT cannot be solved in $O(n^{2 \cos(\pi/7)-o(1)})$ time and $n^{o(1)}$ space.
%\end{theorem}
%\proof  (Sketch) First, one can show that if the desired lower bound exponent $c$ is less than $2$, then one can replace occurrences of $0, 0$ with $0, 1, 0, 0$ in an annotation, and the resulting LP is feasible if the original LP was feasible. Moreover, occurrences of $0, 0, 0$ can be replaced with $0, 1, 0, 0, 1, 0, 0$. This implies that every annotation can be assumed to have at most three consecutive zeroes, and $1, 0$ sequences may be freely inserted between two $0$'s. Hence we can transform the annotations in the theorem statement to those in the general form $1^k (10)^* 0 (10)^* 0 \cdots (10)^* 0$. But one can prove formally that the best lower bound provable over this subset of annotations is the $2 \cos(\pi/7)$ lower bound.\qed
%A similar argument applies to all other annotations found by computer.
We are led to:

\begin{conjecture}\label{nobetter} {\em There is no alternation-trading proof that $\NT[n] \isnotin \DTS[n^c]$ for all $c > 2 \cos(\pi/7)$.}
\end{conjecture}

Proving the conjecture seems currently out of reach. However, we can show:% and we believe that it can be extended further. %It was already believed that alternation-trading proofs of lower bounds for SAT were limited, in that a quadratic time lower bound appeared to be the best that one could do.

\begin{theorem}\label{noquad} There is no alternation-trading proof that $\NT[n] \isnotin \DTS[n^2]$.
\end{theorem}

A proof is in the full version. At a high level, the proof argues that any minimum length proof of a quadratic lower bound could be shortened, giving a contradiction. %If the conjecture is true, it is very useful knowledge in that it forces us to rethink our whole approach. In order to significantly exceed the current best result, it appears we cannot expect to rearrange the existing ingredients---we must find ways to mix in additional complexity-theoretic components. %There are plenty of tools out there, and we believe the search is not over but instead is only beginning.%; the challenge is to find those that can better sharpen our understanding of time lower bounds.

Despite this bad news, the theorem prover did provide enough insight to aid in a new lower bound of $n^{2 \cos(\pi/7)-o(1)}$ on the time-space product of any SAT algorithm.

\begin{theorem} Let $t(n)$ and $s(n)$ be bounded above by polynomials. Any algorithm solving SAT in time $t$ and space $s$ requires $t\cdot s = \Omega(n^{2 \cos(\pi/7)-\eps})$ for all $\eps > 0$.
\end{theorem}

These lower bounds have also been generalized to the QBF problem:

\begin{theorem} For all $k \geq 1$, $\kQBF$ requires $\Omega(n^c)$ time on $n^{o(1)}$ space RAMs, where $c^3/k - c^2 - 2c + k < 0$.
\end{theorem}

\section{Discussion}

We introduced a methodology for reasoning about alternation-trading proofs of lower bounds. It provides a generic means for computers to help us attack lower bound problems, and lets us establish limitations on known techniques.
We now have a better understanding of what these techniques can and cannot do, and a tool for addressing future problems. Previously, the problem of setting parameters to achieve a good lower bound was a highly technical exercise. Our work should facilitate further research: once a new speedup or slowdown lemma is found, one only needs to find the relevant linear programming formulation to begin understanding its power. We conclude with two open-ended problems.\begin{enumerate}

\item {\em Establish tight limitations for alternation-trading proofs.} That is, show that the best possible alternation-trading proofs match those we have provided. Our computer search results have been met with healthy skepticism. It is critical to verify these perceived limitations with formal proof. We have managed to prove non-trivial limitations; it is possible that the ideas in those can be extended.

\item {\em Discover new ingredients to add to the framework.} One possibility is to find new separation results that lead to new contradictions. Another is to find improved Speedup and/or Slowdown Lemmas. The Slowdown Lemmas are the ``blandest'' of the ingredients, in that they are the most elementary (and they relativize). \end{enumerate}

{\small \noindent{\bf Acknowledgements.} I am grateful to my thesis committee for their invaluable feedback on my PhD thesis, which included preliminary results on this work. Thanks to Scott Aaronson for useful discussions about irrelativization, and thanks to the STACS referees for very thoughtful comments.}


\begin{thebibliography}{99}

\bibitem[AKRRV01]{AKRRV} E. Allender, M. Koucky, D. Ronneburger, S. Roy, and V. Vinay. Time-space tradeoffs in the counting hierarchy. In {\em Proc. IEEE Conference on Computational Complexity (CCC)}, 295--302, 2001.

\bibitem[CKS81]{CKS} A. K. Chandra, D. Kozen, and L. J. Stockmeyer. Alternation. {\em JACM} 28(1):114--133, 1981.

\bibitem[Coo88]{Cook} S. A. Cook. Short propositional formulas represent nondeterministic computations. {\em IPL} 26(5): 269-270, 1988.

\bibitem[Der72]{Derman} C. Derman. {\em Finite state Markov decision processes.} Academic Press, 1972.

\bibitem[DvM06]{DvM} S. Diehl and D. van Melkebeek. Time-space lower bounds for the polynomial-time hierarchy on randomized machines. {\em SIAM J. Computing} 36: 563-594, 2006.

\bibitem[DvMW09]{DvM2} S. Diehl, D. van Melkebeek, and R. Williams. An improved time-space lower bound for tautologies. In {\em Proc. of Computing and Combinatorics (COCOON)}, Springer LNCS 5609, 429--438, 2009.

\bibitem[For97]{F} L. Fortnow. Nondeterministic polynomial time versus nondeterministic logarithmic space. In {\em Proc. IEEE Conference on Computational Complexity (CCC)}, 52--60, 1997.

\bibitem[FvM00]{FvM} L. Fortnow and D. van Melkebeek. Time-Space Tradeoffs for Nondeterministic Computation. In {\em Proc. IEEE Conference on Computational Complexity (CCC)}, 2--13, 2000.

\bibitem[FLvMV05]{FLvMV} L. Fortnow, R. Lipton, D. van Melkebeek, and A. Viglas. Time-Space Lower Bounds for Satisfiability. {\em JACM} 52(6):835--865, 2005.

\bibitem[HLMW86]{Halpern} J. Y. Halpern, M. C. Loui, A. R. Meyer, and D. Weise. On Time versus Space III. {\em Mathematical Systems Theory} 19(1):13--28, 1986.

\bibitem[HPV77]{HPV} J. Hopcroft, W. Paul, and L. Valiant. On time versus space. {\em JACM} 24(2):332--337, 1977.

\bibitem[Kan83]{K83} R. Kannan. Alternation and the power of nondeterminism. In {\em Proc. ACM STOC}, 344--346, 1983.

\bibitem[Kan84]{K84} R. Kannan. Towards separating nondeterminism from determinism. {\em Mathematical Systems Theory} 17(1):29--45, 1984.

\bibitem[LV99]{LV} R. J. Lipton and A. Viglas. On the complexity of SAT. In {\em Proc. IEEE FOCS}, 459--464, 1999.

\bibitem[Lou80]{Loui} M. C. Loui. Simulations among multidimensional Turing machines. Ph.D. Thesis, Massachusetts Institute of Technology TR-242, 1980.

\bibitem[MS87]{MS} W. Maass and A. Schorr. Speed-up of Turing machines with one work tape and a two-way input tape. {\em SIAM J. Computing} 16(1):195--202, 1987.

\bibitem[vM04]{VMsurvey} D. van Melkebeek. Time-space lower bounds for NP-complete problems. In {\em Current Trends in Theoretical Computer Science} 265--291, World Scientific, 2004.

\bibitem[vM07]{VMsurvey2} D. van Melkebeek. A survey of lower bounds for satisfiability and related problems. {\em Foundations and Trends in Theoretical Computer Science} 2(3):197--303, 2007.

\bibitem[vMR05]{vMR}  D. van Melkebeek and R. Raz. A time lower bound for satisfiability. {\em TCS} 348(2-3):311--320, 2005.

\bibitem[vMW07]{vMW} D. van Melkebeek and T. Watson. A quantum time-space lower bound for the counting hierarchy. Technical Report 1600, Department of Computer Sciences, University of Wisconsin-Madison, 2007.

\bibitem[Nep70]{Nep} V. Nepomnjascii. Rudimentary predicates and Turing calculations. {\em Soviet Math. Doklady} 11:1462--1465, 1970.

\bibitem[PR81]{PR} W. Paul and R. Reischuk. On time versus space II. {\em JCSS} 22:312--327, 1981.

\bibitem[PPST83]{PPST} W. Paul, N. Pippenger, E. Szemeredi, and W. Trotter. On determinism versus nondeterminism and related problems. In {\em Proc. IEEE FOCS}, 429--438, 1983.

\bibitem[Sch78]{Schnorr} C. Schnorr. Satisfiability is quasilinear complete in NQL. {\em JACM} 25(1):136--145, 1978.

\bibitem[Tou01]{Tourlakis}  I. Tourlakis. Time-space tradeoffs for SAT on nonuniform machines. {\em JCSS} 63(2):268--287, 2001.

\bibitem[Vio09]{Viola} E. Viola. On approximate majority and probabilistic time. {\em Computational Complexity} 18(3):337--375, 2009.

\bibitem[Wil06]{W06} R. Williams. Inductive time-space lower bounds for SAT and related problems. {\em  Computational Complexity} 15:433--470, 2006.

\bibitem[Wil07]{W07} R. Williams. Algorithms and resource requirements for fundamental problems. Ph.D. Thesis, Carnegie Mellon University, CMU-CS-07-147, August 2007.

\bibitem[Wil08]{W08} R. Williams. Time-space tradeoffs for counting NP solutions modulo integers. {\em  Computational Complexity} 17(2):179--219, 2008.

\end{thebibliography}
\end{document}